# Optimization of a fiber Fabry-Perot resonator for low-threshold modulation instability Kerr frequency combs

Germain Bourcier,[1,2,*] Safia Mohand Ousaid,[1] Stephane Balac,[3] Julien Lumeau,[4] Antonin Moreau,[4] Thomas Bunel,[5] Arnaud Mussot,[5] Matteo Conforti,[5] Olivier Llopis,[1] and Arnaud Fernandez,[1]

[1]LAAS-CNRS, Université de Toulouse, CNRS, 7 avenue du Colonel Roche, 31031 Toulouse, France
[2]CNES, 18 Avenue Edouard Belin, F-31401 Toulouse, France
[3]IRMAR, Université de Rennes, CNRS, Campus de Beaulieu, 35042 Rennes, France
[4]Aix Marseille Univ, CNRS, Centrale Med, Institut Fresnel, Marseille, France
[5]University of Lille, CNRS, UMR 8523-PhLAM Physique des Lasers Atomes et Molecules, F-59000, Lille, France
*germain.bourcier@laas.fr

**We report a theoretical and experimental investigation of fiber Fabry-Perot cavities aimed at enhancing Kerr frequency comb generation. The modulation instability (MI) power threshold is derived from the linear stability analysis of a generalized Lugiato-Lefever equation. By combining this analysis with the concepts of power enhancement factor (PEF) and optimal coupling, we predict the ideal manufacturing parameters of fiber Fabry-Perot (FFP) cavities for MI Kerr frequency comb generation. Our findings reveal a distinction between the optimal coupling for modulation instability and that of the cold cavity. Consequently, mirror reflectivity must be adjusted to suit the specific application. We verified the predictions of our theory by measuring the MI power threshold as a function of detuning for three different cavities.**

## 1. INTRODUCTION

In the realm of optical frequency combs (OFC) based on compact comb sources [1–5], fiber Fabry-Perot (FFP) resonators with a Q factor ranging between $10^7$ and $10^8$ emerge as a compelling alternative to ring cavity (RC) resonators. Over the last decade, the latter have garnered significant interest, primarily due to their diverse applications, ranging from spectroscopy [6] and LIDAR [7,8] to telecommunication [9,10] and astronomy[11,12]. They have undergone extensive theoretical and experimental exploration in various forms, including microresonators [3,13–15] and fiber ring cavities [16–18]. Pioneering work by Braje et al. [19], followed by Obrzud et al. [20], marked the initiation of research into the application of FFP for OFC generation. FFP resonators consist of centimeter long fiber sections enclosed by multilayer dielectric mirrors, each a few microns thick (Fig. 1). In addition to their straightforward design and handling, these resonators ensure robust and reproducible coupling, facilitated by FC/PC connectors. They feature ceramic ferrules to simplify coupling, along with facets polishing and coating, employing highly non-linear fiber (HNL) and single-mode fiber (SMF28). OFC generation by Nie et al. [21] utilizing multimode 50 µm core fiber is noteworthy. Nie's findings highlighted losses associated with the diffraction of the fundamental mode in the mirror, underscoring the advantages of multimode fibers in mitigating these effects thus increasing the Q factor to $10^9$. The reported results were obtained in the under-coupling linear regime, where the power reflectivity coefficient ($R$) exceeds the attenuation coefficient ($a$). Notably, the cold cavity was not optimized, particularly regarding intracavity power ($P_{IC}$). A cavity is deemed "hot" when nonlinear effects emerge, leading to a drift of the resonance towards lower frequencies. Modulation instability (MI), is one of the key mechanisms underlying the generation of optical Kerr frequency combs, MI induced combs were recently theoretically [22] and experimentally studied in a single-mode FFP using pulsed pumping at 1550 nm [23]. This paper aims to elucidate FFP cavity coupling, the MI power threshold, and their dependencies on the actual cavity manufacturing parameters.

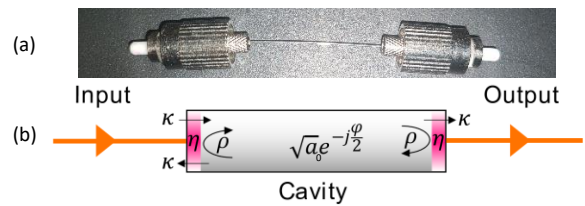

Fig. 1. (a) Fiber Fabry-Perot resonator measuring 7 cm in length. Mirrors are applied to the surface of fiber optic ferrules, which serve as terminations for the cord, functioning as standard fiber connectors, (b) Schematic representation of a Fabry-Perot cavity.

We present an analytical expression for the MI input power threshold, factoring in mirrors' reflectivity and cavity losses. Through this study, we aspire to offer a deeper comprehension of the resonator and manufacturing conditions, fostering the development of more tailored and less energy consuming cavities for applications in both linear regimes and nonlinear OFC generation. The paper is organized as follows: In Sec. 2, we delve into the optical gain in Fabry-Perot resonators and the concept of optimal coupling. Subsequently, in Sec. 3, we derive the MI gain in the good cavity limit and express the MI input power threshold as a function of mirror reflectivity, losses and detuning. Finally, in Sec. 4, we present experimental results on MI threshold.

## 2. COLD CAVITY RESONANCE

### A. Fabry-Perot resonator model

Fiber Fabry-Perot (FFP) configuration comprises a fiber segment of length $L$, terminated by two identical dielectric mirrors characterized by a reflectivity coefficient $\rho$ and a transmissivity coefficient $\kappa$. We introduce the transfer function that characterizes resonances spaced at fixed intervals in frequency, determined by the free spectral range, $FSR = c/(2n_g L)$ with $c$ representing the speed of light, and $n_g$ being the group refractive index. Dimensionless linear propagation losses denoted by $a_0$ correspond to losses in a complete roundtrip, and they are related to the imaginary part of the propagation constant $\alpha_0/2$ (m$^{-1}$) as $a_0 = e^{-\alpha_0 L}$. Additional intracavity losses, encompassing mirror absorption and diffraction, are accounted for by a parameter $\eta$. Intrinsic losses are then defined by $a = \eta^2 a_0$. We first describe the resonator in the linear regime and with a single-mode fiber. The transfer function of the FFP resonator can then be expressed in terms of the roundtrip phase $\varphi$ as follows:

$$|T(\varphi)|^2 = \frac{\kappa^4 a}{1 + (a\rho^2)^2 - 2a\rho^2 \cos(\varphi)} \quad (1)$$

where $\kappa^2 + \rho^2 = 1$. The transmission at resonance is $T_{max} = |T(\varphi = 0)|^2$ (see Fig. 2).

### B. Power enhancement factor and optimal coupling

While achieving maximal transmission is crucial, the primary objective in OFC generation is to reach a significant intracavity power $P_{IC}$. To obtain intra-cavity power build-up at resonance, a high-quality (Q) resonator is employed. The intracavity power enhancement factor (PEF) is a function of $a$ and $\rho$, defined as the ratio of intracavity power to cavity input power as:

$$\text{PEF} = \frac{1 - \rho^2}{(1 - a\rho^2)^2} \quad (2)$$

PEF reaches its maximum for an optimal reflectivity $R_{opt} = \rho_{opt}^2$ when $\rho_{opt}^2 = (2a - 1)/a$, which can be approximated at the first order to $\rho_{opt}^2 = a$ in the vicinity of $a = 1$. The coupling characteristics of FFP cavity are different from a ring cavity, especially in the case of equal mirrors. Indeed, only under-coupling is possible in this case [24]. All the conditions associated to critical coupling (maximum PEF, total absorption and 0 dB transmission) cannot be met for the same reflectivity value. Therefore, we define the optimal coupling of a FFP resonator as the condition of maximum PEF. At optimal coupling, the resonator's transmission is equal to $\frac{1}{4a} \approx -6\,\text{dB}$. By varying $a$ and $\rho$, the coupling regime changes from over-coupling to under-coupling with respect to optimal coupling. Over-coupling arises when reflectivity is too low, resulting in excessive light exiting the cavity. Conversely, under-coupling occurs when reflectivity is too high, leading to insufficient light entering the cavity, signifying low transmissivity of the entering mirror. In Figure 2, the transmission at resonance $T_{max}$ and PEF are plotted as a function of power reflectivity $R$ with $a = 0.9985$ (retrieved from our spectro-RF measurement setup [25] with 7.35 cm long single mode fiber FFP). We observe $R_{opt}$ corresponding to the maximal power enhancement factor $\text{PEF}_{max}$ when $T_{max}$ is approximately -6 dB. This singular property was confirmed experimentally (Fig. 2. (a)). For $R < R_{opt}$, transmission exceeds -6 dB, indicating over-coupling. Conversely, under-coupling is indicated when $T_{max}$ is less than -6 dB with $R > R_{opt}$.

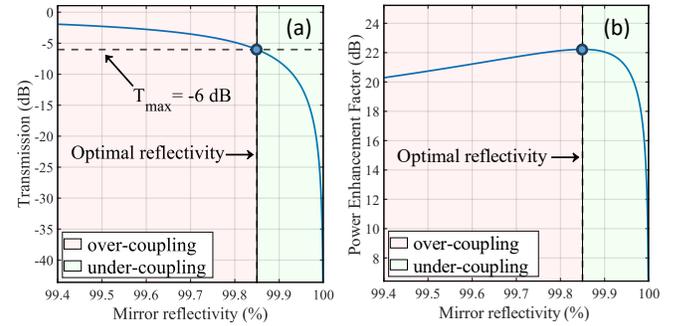

Fig. 2. (a) Transmission at resonance of a FFP resonator (b) PEF, plotted as a function of mirror power reflectivity $R$ with $a = 99.85\,\%$. An optimal coupling (blue circle) is obtained and confirm experimentally with $R = 0.9985$.

## 3. HOT CAVITY RESONANCE

The designation "hot" is assigned to the cavity once the average power within it reaches levels sufficient to induce alterations in the fiber's refractive index, primarily due to thermal or third-order nonlinear effects. Our discussion will specifically emphasize nonlinear effects, particularly the optical Kerr effect. The parameters $a$ and $\rho$ in a hot cavity are defined in the same way as in a cold cavity. The refractive index varies with optical intensity. Apart from the index change, the optical Kerr effect is characterized by four-wave mixing (FWM). FWM manifests itself in a degenerate manner at specific frequencies, driven by dispersive instabilities, also known as temporal modulation instability.

### A. Modulation instability

The modulation instability phenomenon has been recently comprehensively elucidated beyond the mean field limit [22]. In the present study, we opt for the mean field approximation, often referred to as the good-cavity approximation, yielding a specific Lugiato-Lefever equation (LLE) for Fabry-Perot resonators (FP-LLE). Initially derived from Maxwell-Bloch equations [26], the FP-LLE has also been established from coupled mode theory [27] and nonlinear Schrödinger (NLS) equations [22]. Here to align with experimental results, we consider intrinsic losses $a$ yielding the following FP-LLE.

$$t_R \frac{\partial \psi}{\partial \tau} = -(1-a\rho^2)\psi - i\delta\psi + \kappa E_{in}$$

$$+ 2L\left[-i\frac{\beta_2}{2}\frac{\partial^2}{\partial t^2} + i\gamma|\psi|^2 + i\gamma\frac{2}{t_R}\int_{-t_R/2}^{t_R/2}|\psi|^2 dt\right]\psi \quad (3)$$

where $\psi(\tau, t)$ represents the electric field envelope within the cavity, $t_R$ is the roundtrip time, $t$ is the fast time within one cavity roundtrip ($t \in [-t_{R/2}, t_{R/2}]$) and $\tau$ is the slow time over cavity roundtrips. $\beta_2$ is the group velocity dispersion coefficient, $\gamma$ is the nonlinear coefficient and $\delta$ is the cavity detuning. The input and intracavity powers, respectively $P_{IN}$ and $P_{IC}$, of the steady state satisfying the equation:

$$\frac{P_{IC}}{P_{IN}} = \frac{1-\rho^2}{(1-a\rho^2)^2 + (6L\gamma P_{IC} - \delta)^2} \quad (4)$$

From Eq. (4) we see that PEF is the maximum of the ratio $P_{IC}/P_{IN}$ obtained from FP-LLE is identical to Eq. (3). We have conducted a linear stability analysis around the stationary state following the method used in [26,28,29]. Derivation method adapted to FFP can also be found in [22]. This approach enabled us to compute the MI parametric gain spectrum for given intra-cavity power and detuning. The MI gain exhibits two lobes symmetrically positioned around the pump frequency of the incoming field. An optical power gain is observed corresponding to these MI lobes, with the maximum gain $g_{max}$ that can be expressed in terms of the intra-cavity power and losses:

$$g_{max} = -\frac{1-a\rho^2}{2L} + \gamma P_{IC} \quad (5)$$

where $\gamma = 2\pi n_2/(\lambda A_{eff})$ and $A_{eff}$ refers to the effective area of the fiber. The intracavity power required to surpass losses and attain the maximum gain $g_{max} > 0$, reads as:

$$P_{IC,th} = \frac{1-a\rho^2}{2L\gamma} \quad (6)$$

Beyond this MI power threshold $P_{IC,th}$ the generation of photons occurs through degenerate FWM from the pump frequency to frequencies within the MI lobes.

### B. MI power threshold

Considering both $P_{IC}/P_{IN}$ (Eq. 4) and $P_{IC,th}$ (Eq. 6), we can compute the necessary injected power in order to reach MI threshold ($P_{IN,th}$) and appreciate its dependence on the cavity detuning $\delta$ and the reflectivity $\rho$ according to:

$$P_{IN,th} = \frac{1-a\rho^2}{2L\gamma(1-\rho^2)}[(1-a\rho^2)^2 + (3(1-a\rho^2) - \delta)^2] \quad (7)$$

Input MI threshold (Eq. 7) is plotted in Fig. 3 as a function of reflectivity and detuning. We can appreciate graphically that $P_{IN,th}$ reaches a unique minimum value $P_{IN,th,min}$ labeled by a red spot which is given by:

$$P_{IN,th,min} = \frac{27(1-a)^2 a}{8L\gamma} \quad (8)$$

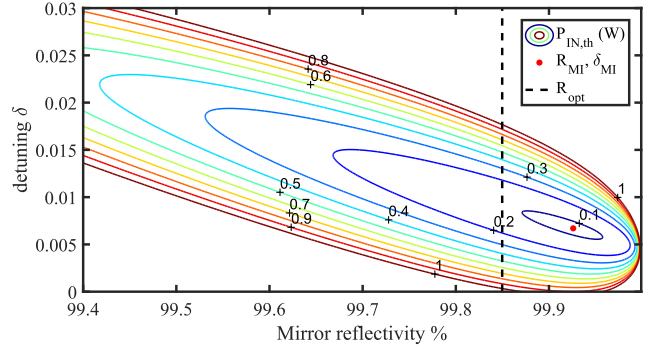

Fig. 3. Contour plot of input MI power threshold as a function of mirror reflectivity and detuning with intrinsic losses $a = 0.9985$. We show the cold cavity optimal reflectivity obtained with PEF (black dashed line) and the MI optimal reflectivity $R_{MI} = 99.925\%$ and detuning $\delta = 0.0068$ (red point).

The corresponding reflectivity $R_{MI} = \rho_{MI}^2$ and detuning $\delta_{MI}$ are expressed analytically as a function of intrinsic losses:

$$\rho_{MI}^2 = \frac{3a-1}{2a} \quad (9)$$

$$\delta_{MI} = \frac{9}{2}(1-a) \quad (10)$$

Hence, a new optimal reflectivity $R_{MI}$ is observed for the minimization of $P_{IN,th}$. We can appreciate graphically (Fig. 3) and analytically that $R_{MI} > R_{opt}$ for a passive cavity ($a < 1$). The obtained $P_{IN,th,min}$ has to be compared to the minimum power threshold when taking $R_{opt}$ which is $P_{IN,th,R_{opt}} = 32(1-a)^2 a/8L\gamma$. This optimal reflectivity is shifted towards higher reflectivity in order to minimize $P_{IN,th,min}$. The reduction in power is $5/32 = 15.625\%$. This result highlights the necessity of under-coupling the resonator to optimize the generation of MI in FFP. This under-coupling is crucial, even though the resonator will not be optimized in terms of linear transmission. In this scenario, we achieved an optimum transmission loss of -11 dB (compared to -6 dB to maximize PEF), concurrently increasing both the Q factor and finesse. Within the scope of FFP resonator with a Q factor generally approaching $10^8$ this enhancement is a significant advantage due the required high input power. It is also important to note that the design of under-coupled FFP dedicated to lower the MI threshold will not be compatible and optimized for linear applications such as optoelectronic oscillator whose cold cavity performance is compulsory. Regardless of the reflectivity, the best detuning always corresponds to the maximum of the tilted resonance reaching the power threshold (Fig. 4(a)).

### 4. MI TRESHOLD MEASUREMENT

Since controlling losses during the manufacturing process is challenging, comparing resonators with varying reflectivity but similar intracavity losses are difficult to realize. However, we can readily conduct a study based on the variation of detuning. This is the approach we propose in this experimental section. We select three resonators of different fiber types, each with nearly identical losses and reflectivity, resulting in a finesse $F$ around 500.

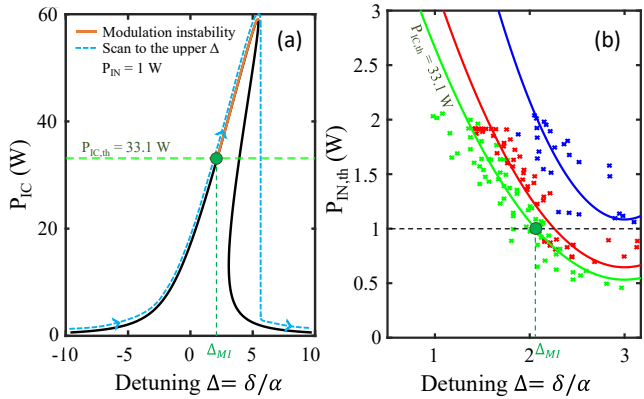

Fig. 4. (a) Theoretical nonlinear transfer function of a LEAF resonator of 7 cm with $R = 0.9984$ and $a = 0.995$ for a scan to the upper $\Delta$ (blue dashed line). MI area is highlighted in orange. (b) MI input power threshold as a function of detuning for three different fiber resonators with a close couple $(R, a)$. Solid lines are theory [Eq. (7)] while crosses show experimental data. Green: LEAF fiber as in (a); red: HNL fiber; blue: SMF28 fiber.

Mirrors are designed by depositing 9 pairs of alternated high ($n_H$=2.2) and low ($n_L$=1.47) refractive index quarterwave layers achieving a reflectivity of 99.84% at 1550 nm and a total thickness of 4.2 µm. Fig. 4(a) illustrates the method used to link input power threshold and detuning with large effective area fiber (LEAF) ($A_{eff} = 72\ \mu m^2$) resonator features. Starting from a negative detuning and by increasing the detuning, intracavity power follows the upper branch to the tilted resonance peak and then jumps to the lower branch (blue dashed line in Fig.4(a)). MI is observable when $P_{IC}$ exceeds $P_{IC,th}$ (green circle in Fig. 4). By experimentally detecting the beginning of MI we can determine the normalized detuning $\Delta_{MI} = \delta_{MI}/\alpha$, where $\alpha = 1 - a\rho^2$, for which the pump power equals $P_{IN,th}$. By increasing the pump power from 0 to 2 W, and by repeating the detuning scans, we obtain the detuning values for which MI appears as a function of the pump power (Fig. 4(b)). The theoretical predictions are plotted in Fig. 4(b) (blue, green and red solid lines) using Eq. (7). The three resonators have similar reflectivity and losses but the HNL one ($A_{eff} = 12\ \mu m^2$; red line and crosses in Fig. 4(b)) has additional insertion losses because the upstream system is single-mode (SMF28) with a higher effective area $A_{eff} = 85\ \mu m^2$. Insertion losses elucidate why the input power threshold $P_{IN,th}$ is higher than expected with such a highly nonlinear fiber. The experimental measurements are in perfect agreement with our theoretical prediction.

## 5. CONCLUSION

We analyzed the impact of mirror reflectivity on the performance of a FFP resonator. We have shown that optimizing the resonator for nonlinear operation result in a different set of parameters than for linear regime. We have revealed that employing an under-coupled cavity by a careful selection of the reflectivity can significantly reduce by 15.6% the pump power required to induce a modulation instability comb. This result is particularly noteworthy as FFP systems require intense pumping for successful comb generation; Hence, reducing this threshold marks a substantial advancement in comb generation. Moreover, we have shown that the laser detuning can also be leveraged to optimize MI generation, yielding results that closely reproduce our experimental results. Collectively, our findings deepen the comprehension of FFP design prerequisites essential for the generation of Kerr frequency combs.

**Funding.** The present research was supported by the French Ministry of Armed Forces – Agence de l'Innovation Défense (AID) and the national space center (CNES).

**Disclosures.** The authors declare no conflicts of interest.

**Data availability**. Data underlying the results presented in this paper are not publicly available at this time but may be obtained from the authors upon reasonable request.

## References

1. A. Pasquazi, M. Peccianti, L. Razzari, D. J. Moss, S. Coen, M. Erkintalo, Y. K. Chembo, T. Hansson, S. Wabnitz, P. Del'Haye, X. Xue, A. M. Weiner, and R. Morandotti, Physics Reports **729**, 1 (2018).
2. T. J. Kippenberg, A. L. Gaeta, M. Lipson, and M. L. Gorodetsky, Science **361**, eaan8083 (2018).
3. T. Fortier and E. Baumann, Commun Phys **2**, 153 (2019).
4. W. Wang, L. Wang, and W. Zhang, Adv. Photon. **2**, 1 (2020).
5. C. Silvestri, X. Qi, T. Taimre, K. Bertling, A. D. Rakic, APL Photonics **8,** (2023).
6. M.-G. Suh, Q.-F. Yang, K. Y. Yang, X. Yi, and K. J. Vahala, Science **354**, 600 (2016).
7. M.-G. Suh and K. J. Vahala, Science **359**, 884 (2018).
8. J. Riemensberger, A. Lukashchuk, M. Karpov, W. Weng, E. Lucas, J. Liu, and T. J. Kippenberg, Nature **581**, 164 (2020).
9. P. Marin-Palomo, J. N. Kemal, M. Karpov, A. Kordts, J. Pfeifle, M. H. P. Pfeiffer, P. Trocha, S. Wolf, V. Brasch, M. H. Anderson, R. Rosenberger, K. Vijayan, W. Freude, T. J. Kippenberg, and C. Koos, Nature **546**, 274 (2017).
10. B. Corcoran, M. Tan, X. Xu, A. Boes, J. Wu, T. G. Nguyen, S. T. Chu, B. E. Little, R. Morandotti, A. Mitchell, and D. J. Moss, Nat Commun **11**, 2568 (2020).
11. E. Obrzud, M. Rainer, A. Harutyunyan, M. H. Anderson, J. Liu, M. Geiselmann, B. Chazelas, S. Kundermann, S. Lecomte, M. Cecconi, A. Ghedina, E. Molinari, F. Pepe, F. Wildi, F. Bouchy, T. J. Kippenberg, and T. Herr, Nature Photon **13**, 31 (2019).
12. M.-G. Suh, X. Yi, Y.-H. Lai, S. Leifer, I. S. Grudinin, G. Vasisht, E. C. Martin, M. P. Fitzgerald, G. Doppmann, J. Wang, D. Mawet, S. B. Papp, S. A. Diddams, C. Beichman, and K. Vahala, Nature Photon **13**, 25 (2019).
13. T. Herr, V. Brasch, J. D. Jost, C. Y. Wang, N. M. Kondratiev, M. L. Gorodetsky, and T. J. Kippenberg, Nature Photon **8**, 145 (2014).
14. J. Liu, E. Lucas, A. S. Raja, J. He, J. Riemensberger, R. N. Wang, M. Karpov, H. Guo, R. Bouchand, and T. J. Kippenberg, Nat. Photonics **14**, 486 (2020).
15. E. Lucas, P. Brochard, R. Bouchand, S. Schilt, T. Südmeyer, and T. J. Kippenberg, Nat Commun **11**, 374 (2020).
16. F. Leo, S. Coen, P. Kockaert, S.-P. Gorza, P. Emplit, and M. Haelterman, Nature Photon **4**, 471 (2010).
17. N. Englebert, F. De Lucia, P. Parra-Rivas, C. M. Arabí, P.-J. Sazio, S.-P. Gorza, and F. Leo, Nat. Photon. **15**, 857 (2021).
18. N. Englebert, S.-P. Gorza, and F. Leo, Nat. Photon. **15**, 536 (2021).
19. D. Braje, L. Hollberg, and S. Diddams, Phys. Rev. Lett. **102**, 193902 (2009).
20. E. Obrzud, S. Lecomte, and T. Herr, Nature Photon **11**, 600 (2017).
21. M. Nie, K. Jia, Y. Xie, S. Zhu, Z. Xie, and S.-W. Huang, Nat Commun **13**, 6395 (2022).
22. Z. Ziani, T. Bunel, A. M. Perego, A. Mussot, and M. Conforti, Phys. Rev. A **109**, 013507 (2024).
23. T. Bunel, M. Conforti, Z. Ziani, J. Lumeau, A. Moreau, A. Fernandez, O. Llopis, J. Roul, A. M. Perego, K. K. Y. Wong, and A. Mussot, Opt. Lett. **48**, 275 (2023).
24. J. Musgrave, S.-W. Huang, and M. Nie, APL Photonics **8**, (2023).
25. Z. Abdallah, Y. G. Boucher, A. Fernandez, S. Balac, and O. Llopis, Sci Rep **6**, 27208 (2016).
26. D. C. Cole, A. Gatti, S. B. Papp, F. Prati, and L. Lugiato, Phys. Rev. A **98**, 013831 (2018).
27. Z. Xiao, K. Wu, T. Li, and J. Chen, Opt. Express **28**, 14933 (2020).
28. W. J. Firth, J. B. Geddes, N. J. Karst, and G.-L. Oppo, Phys. Rev. A **103**, 023510 (2021).
29. M. Conforti, A. Mussot, A. Kudlinski, and S. Trillo, Opt. Lett. **39**, 4200 (2014).

## References

1. A. Pasquazi, M. Peccianti, L. Razzari, D. J. Moss, S. Coen, M. Erkintalo, Y. K. Chembo, T. Hansson, S. Wabnitz, P. Del'Haye, X. Xue, A. M. Weiner, and R. Morandotti, "Micro-combs: A novel generation of optical sources," Physics Reports **729**, 1 (2018).
2. T. J. Kippenberg, A. L. Gaeta, M. Lipson, and M. L. Gorodetsky, "Dissipative Kerr solitons in optical microresonators," Science **361**, eaan8083 (2018).
3. T. Fortier and E. Baumann, "20 years of developments in optical frequency comb technology and applications," Commun Phys **2**, 153 (2019).
4. W. Wang, L. Wang, and W. Zhang, "Advances in soliton microcomb generation," Adv. Photon. **2**, 1 (2020).
5. C. Silvestri, X. Qi, T. Taimre, K. Bertling, A. D. Rakic, "Frequency combs in quantum cascade lasers: An overview of modeling and experiments," APL Photonics **8**, (2023)
6. M.-G. Suh, Q.-F. Yang, K. Y. Yang, X. Yi, and K. J. Vahala, "Microresonator soliton dual-comb spectroscopy," Science **354**, 600 (2016).
7. M.-G. Suh and K. J. Vahala, "Soliton microcomb range measurement," Science **359**, 884 (2018).
8. J. Riemensberger, A. Lukashchuk, M. Karpov, W. Weng, E. Lucas, J. Liu, and T. J. Kippenberg, "Massively parallel coherent laser ranging using a soliton microcomb," Nature **581**, 164 (2020).
9. P. Marin-Palomo, J. N. Kemal, M. Karpov, A. Kordts, J. Pfeifle, M. H. P. Pfeiffer, P. Trocha, S. Wolf, V. Brasch, M. H. Anderson, R. Rosenberger, K. Vijayan, W. Freude, T. J. Kippenberg, and C. Koos, "Microresonator-based solitons for massively parallel coherent optical communications," Nature **546**, 274 (2017).
10. B. Corcoran, M. Tan, X. Xu, A. Boes, J. Wu, T. G. Nguyen, S. T. Chu, B. E. Little, R. Morandotti, A. Mitchell, and D. J. Moss, "Ultra-dense optical data transmission over standard fibre with a single chip source," Nat Commun **11**, 2568 (2020).
11. E. Obrzud, M. Rainer, A. Harutyunyan, M. H. Anderson, J. Liu, M. Geiselmann, B. Chazelas, S. Kundermann, S. Lecomte, M. Cecconi, A. Ghedina, E. Molinari, F. Pepe, F. Wildi, F. Bouchy, T. J. Kippenberg, and T. Herr, "A microphotonic astrocomb," Nature Photon **13**, 31 (2019).
12. M.-G. Suh, X. Yi, Y.-H. Lai, S. Leifer, I. S. Grudinin, G. Vasisht, E. C. Martin, M. P. Fitzgerald, G. Doppmann, J. Wang, D. Mawet, S. B. Papp, S. A. Diddams, C. Beichman, and K. Vahala, "Searching for Exoplanets Using a Microresonator Astrocomb," Nature Photon **13**, 25 (2019).
13. T. Herr, V. Brasch, J. D. Jost, C. Y. Wang, N. M. Kondratiev, M. L. Gorodetsky, and T. J. Kippenberg, "Temporal solitons in optical microresonators," Nature Photon **8**, 145 (2014).
14. J. Liu, E. Lucas, A. S. Raja, J. He, J. Riemensberger, R. N. Wang, M. Karpov, H. Guo, R. Bouchand, and T. J. Kippenberg, "Photonic microwave generation in the X- and K-band using integrated soliton microcombs," Nat. Photonics **14**, 486 (2020).
15. E. Lucas, P. Brochard, R. Bouchand, S. Schilt, T. Südmeyer, and T. J. Kippenberg, "Ultralow-noise photonic microwave synthesis using a soliton microcomb-based transfer oscillator," Nat Commun **11**, 374 (2020).
16. F. Leo, S. Coen, P. Kockaert, S.-P. Gorza, P. Emplit, and M. Haelterman, "Temporal cavity solitons in one-dimensional Kerr media as bits in an all-optical buffer," Nature Photon **4**, 471 (2010).
17. N. Englebert, F. De Lucia, P. Parra-Rivas, C. M. Arabí, P.-J. Sazio, S.-P. Gorza, and F. Leo, "Parametrically driven Kerr cavity solitons," Nat. Photon. **15**, 857 (2021).
18. N. Englebert, S.-P. Gorza, and F. Leo, "Temporal Solitons in a Coherently Driven Active Resonator," Nat. Photon. **15**, 536 (2021).
19. D. Braje, L. Hollberg, and S. Diddams, "Brillouin-Enhanced Hyperparametric Generation of an Optical Frequency Comb in a Monolithic Highly Nonlinear Fiber Cavity Pumped by a cw Laser," Phys. Rev. Lett. **102**, 193902 (2009).
20. E. Obrzud, S. Lecomte, and T. Herr, "Temporal solitons in microresonators driven by optical pulses," Nature Photon **11**, 600 (2017).
21. M. Nie, K. Jia, Y. Xie, S. Zhu, Z. Xie, and S.-W. Huang, "Synthesized spatiotemporal mode-locking and photonic flywheel in multimode mesoresonators," Nat Commun **13**, 6395 (2022).
22. Z. Ziani, T. Bunel, A. M. Perego, A. Mussot, and M. Conforti, "Theory of modulation instability in Kerr Fabry-Perot resonators beyond the mean-field limit," Phys. Rev. A **109**, 013507 (2024).
23. T. Bunel, M. Conforti, Z. Ziani, J. Lumeau, A. Moreau, A. Fernandez, O. Llopis, J. Roul, A. M. Perego, K. K. Y. Wong, and A. Mussot, "Observation of modulation instability Kerr frequency combs in a fiber Fabry-Pérot resonator," Opt. Lett. **48**, 275 (2023).
24. J. Musgrave, S.-W. Huang, and M. Nie, "Microcombs in fiber Fabry-Pérot cavities," APL Photonics **8**, (2023).
25. Z. Abdallah, Y. G. Boucher, A. Fernandez, S. Balac, and O. Llopis, "Radio frequency spectral characterization and model parameters extraction of high Q optical resonators," Sci Rep **6**, 27208 (2016).
26. D. C. Cole, A. Gatti, S. B. Papp, F. Prati, and L. Lugiato, "Theory of Kerr frequency combs in Fabry-Perot resonators," Phys. Rev. A **98**, 013831 (2018).
27. Z. Xiao, K. Wu, T. Li, and J. Chen, "Deterministic single-soliton generation in a graphene-FP microresonator," Opt. Express **28**, 14933 (2020).
28. W. J. Firth, J. B. Geddes, N. J. Karst, and G.-L. Oppo, "Analytic instability thresholds in folded Kerr resonators of arbitrary finesse," Phys. Rev. A **103**, 023510 (2021).
29. M. Conforti, A. Mussot, A. Kudlinski, and S. Trillo, "Modulational instability in dispersion oscillating fiber ring cavities," Opt. Lett. **39**, 4200 (2014).